\newif\ifpreprint
\preprintfalse % Enable for double column reprint

\ifpreprint
\documentclass[journal=jctcce,manuscript=article]{achemso}
\else
\documentclass[journal=jctcce,manuscript=article,layout=twocolumn]{achemso}
\fi
\usepackage{amssymb}

\usepackage{graphicx}% Include figure files
\usepackage{dcolumn}% Align table columns on decimal point
\usepackage{bm}% bold math

\usepackage{lipsum}
\usepackage[table,xcdraw]{xcolor}
\usepackage{multirow}
\usepackage{soul}
\usepackage{simpler-wick}
\usepackage{chemformula}
\usepackage{xr}
\usepackage{hhline}
\usepackage{braket}
\usepackage[]{cleveref}% add hypertext capabilities
\usepackage{listings}
\usepackage{booktabs}
\usepackage{array}
\usepackage{siunitx}
\usepackage{threeparttable}
\usepackage{xcolor}
\usepackage{xspace}
\usepackage[version=4]{mhchem}
\usepackage{anyfontsize}
\usepackage{pdfpages}
\fontsize{9}{11}\selectfont

\crefformat{equation}{#2eq~#1#3}

\newcommand{\adj}{^{\dagger}}

\usepackage[normalem]{ulem}

\lstdefinestyle{mystyle}{
    backgroundcolor=\color{backcolour},   
    commentstyle=\color{codegreen},
    keywordstyle=\color{magenta},
    numberstyle=\tiny\color{codegray},
    stringstyle=\color{codepurple},
    basicstyle=\ttfamily\footnotesize,
    breakatwhitespace=false,         
    breaklines=true,                 
    captionpos=b,                    
    keepspaces=true,                 
    numbers=left,                    
    numbersep=5pt,                  
    showspaces=false,                
    showstringspaces=false,
    showtabs=false,                  
    tabsize=2
}

\ifpreprint
\else

\makeatletter
\let\l@addto@macro\relax
\makeatother
\usepackage[fontsize=9pt]{scrextend}
\fi

\title{Multireference equation-of-motion driven similarity renormalization group for X-ray photoelectron spectra}

\author{Shuhang Li}
\affiliation{Department of Chemistry and Cherry Emerson Center for Scientific Computation, Emory University, Atlanta, Georgia 30322, United States}
\author{Zijun Zhao}
\affiliation{Department of Chemistry and Cherry
Emerson Center for Scientific Computation, Emory
University, Atlanta, Georgia 30322, United States}
\author{Francesco A. Evangelista}
\email{francesco.evangelista@emory.edu}
\affiliation{Department of Chemistry and Cherry Emerson Center for Scientific Computation, Emory University, Atlanta, Georgia 30322, United States}

\begin{document}

\begin{tocentry}
\includegraphics[width=3.25in]{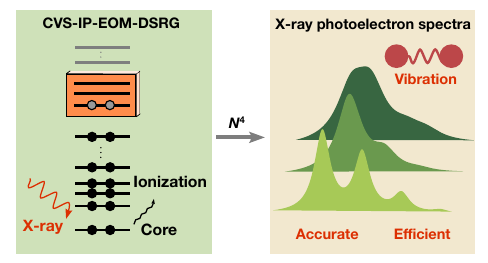}
\end{tocentry}

\begin{abstract}
We formulate and implement the core-valence separated multireference equation-of-motion driven similarity renormalization group method (CVS-IP-EOM-DSRG) for simulating X-ray photoelectron spectra (XPS) of strongly correlated molecular systems.
This method is numerically robust and computationally efficient, delivering accurate core-ionization energies with $\mathcal{O}(N^4)$ scaling relative to basis set size $N$ in the EOM step.
To ensure rigorous core intensivity, we propose a simple modification of the ground-state MR-DSRG formalism.
We develop and compare three variants of the theory based on different approximations of the effective Hamiltonian: two derived from low-order perturbative methods (DSRG-MRPT2 and DSRG-MRPT3), and one from a non-perturbative scheme truncated to 1- and 2-body operators [MR-LDSRG(2)]. 
We benchmark the CVS-IP-EOM-DSRG methods by computing vertical core-ionization energies for a representative molecular test set and comparing results against established single-reference and multireference methods.
To demonstrate the applicability of CVS-IP-EOM-DSRG to strongly correlated systems, we compute the potential energy curves and vibrationally resolved XPS of \ce{N2} and \ce{CO} and the XPS of ozone.
Comparison with experimental data and other high-level theoretical results shows that all three CVS-IP-EOM-DSRG variants accurately predict vertical ionization energies, but only DSRG-MRPT3 and MR-LDSRG(2) levels of theory reliably capture the full dissociation behavior and reproduce the experimental vibrational structure.
\end{abstract}

\maketitle

\section{Introduction}

Time-resolved X-ray photoelectron spectroscopy (TR-XPS) provides element- and site-specific access to ultrafast electronic dynamics.\cite{siefermann.2014.10.1021/jz501264x,neville.2018.10.1103/PhysRevLett.120.243001,brausse.2018.10.1103/PhysRevA.97.043429,inhester.2019.10.1021/acs.jpclett.9b02370,vidal.2020.10.1039/C9CP03695Da,mayer.2022.10.1038/s41467-021-27908-y} By tracking core-level ionizations, TR-XPS reports on oxidation state, local charge redistribution, and spin changes during photochemical and catalytic processes.\cite{neppl.2016.10.1039/C6FD00125D, brausse.2018.10.1103/PhysRevA.97.043429a, dendzik.2020.10.1103/PhysRevLett.125.096401, roth.2021.10.1038/s41467-021-21454-3a, shavorskiy.2021.10.1021/acsami.1c13590, mayer.2022.10.1038/s41467-021-27908-ya, gabalski.2023.10.1021/acs.jpclett.3c01447, gaba.2024.10.1039/D4CP00801Da}
Theory in support of experimental analysis must deliver core-ionization energies and reliable relative intensities along nonequilibrium geometries, often in the presence of strong static correlation and near-degeneracies.\cite{zanchet.2019.10.1021/acs.jpcc.9b08378}

Electronic-structure strategies for core ionization fall into two classes. The first consists of approaches that optimize orbitals, and optionally variational parameters, for each core-ionized state.
This category includes the delta-Hartree--Fock method,\cite{bagus.1965.10.1103/PhysRev.139.A619,navesdebrito.1991.10.1063/1.460898,besley.2009.10.1063/1.3092928} delta-density-functional-theory,\cite{besley.2009.10.1063/1.3092928, triguero.1999.10.1016/S0368-20489900008-0, verma.2016.10.1021/acs.jctc.5b00817, hait.2020.10.1021/acs.jpclett.9b03661, cunha.2022.10.1021/acs.jpclett.2c00578} delta-restricted-active-space self-consistent-field ($\Delta$RASSCF),\cite{jensen.1987.10.1063/1.453590, agren.1993.10.1016/0301-01049380105-I, couto.2020.10.1039/D0CP02207A, lindblad.2020.10.1103/PhysRevLett.124.203001, ghosh.2023.10.1021/acs.jpclett.3c00611, sen.2024.10.1021/acs.jpcc.4c01750} the static exchange (STEX) method,\cite{agren.1994.10.1016/0009-26149400318-1, ekstrom.2006.10.1103/PhysRevA.73.022501} nonorthogonal configuration interaction singles (NOCIS),\cite{oosterbaan.2018.10.1063/1.5023051} and recently developed approaches based on the generalized-active-space self-consistent field (GASSCF) framework.\cite{ma.2011.10.1063/1.3611401, huang.2022.10.1021/acs.jctc.1c00884,huang.2024.10.1021/acs.jctc.4c00835}
These methods capture large core-hole relaxation but incur high cost when many states or geometries are needed; moreover, evaluation of properties is complicated by the lack of orthogonality of the underlying basis set.\cite{burton.2021.10.1063/5.0045442, burton.2022.10.1063/5.0122094}

A second category of methods includes response methods such as the equation-of-motion coupled-cluster theory (EOM-CC)\cite{stanton.1993.10.1063/1.464746, krylov.2008.10.1146/annurev.physchem.59.032607.093602, bartlett.2012.10.1002/wcms.76, coriani.2015.10.1063/1.4935712, vidal.2019.10.1021/acs.jctc.9b00039, liu.2019.10.1021/acs.jctc.8b01160, vidal.2020.10.1039/C9CP03695D, vidal.2020.10.1021/acs.jpclett.0c02027, thielen.2021.10.1063/5.0047134} and the algebraic diagrammatic construction (ADC),\cite{schirmer.1982.10.1103/PhysRevA.26.2395, schirmer.1983.10.1103/PhysRevA.28.1237a, schirmer.1991.10.1103/PhysRevA.43.4647, mertins.1996.10.1103/PhysRevA.53.2140, schirmer.2001.10.1063/1.1418437, thiel.2003.10.1063/1.1584658, schirmer.2004.10.1063/1.1752875, wenzel.2014.10.1002/jcc.23703a, wenzel.2015.10.1063/1.4921841} are particularly well suited for computing many excited states starting from a correlated ground state.
This feature makes them very attractive for predicting pump-probe signals. However, methods in this second category lack explicit orbital relaxation; for core holes, this typically forces the inclusion of triples or transition-potential constructs that compromise ground-state accuracy.\cite{stanton.1993.10.1063/1.464746,schirmer.1982.10.1103/PhysRevA.26.2395,liu.2019.10.1021/acs.jctc.8b01160,simons.2021.10.1063/5.0036631}
To obviate this limitation, Simons and Matthews proposed the transition potential CC (TP-CC) theory, which can accurately describe core-hole states at a purely singles and doubles level.\cite{simons.2021.10.1063/5.0036631, simons.2023.10.1080/00268976.2022.2088421} 
Moreover, EOM and ADC methods have been, for the most part, formulated on the assumption that a single mean-field state dominates the correlated ground state; such an assumption fails to capture the distinctive features of many open-shell systems and transient species, motivating multireference (MR) formulations such as MR-ADC,\cite{sokolov.2018.10.1063/1.5055380, chatterjee.2019.10.1021/acs.jctc.9b00528, chatterjee.2020.10.1021/acs.jctc.0c00778, mazin.2021.10.1021/acs.jctc.1c00684, demoura.2022.10.1039/D1CP05476G, mazin.2023.10.1021/acs.jctc.3c00477, demoura.2024.10.1021/acs.jpca.4c03161} for simulating various spectroscopic processes including XPS.\cite{demoura.2022.10.1039/D1CP05476G}

In this work, we build on a Hermitian MR equation-of-motion framework derived from the multireference driven similarity renormalization group (DSRG),\cite{li.2015.10.1021/acs.jctc.5b00134, li.2016.10.1063/1.4947218, li.2017.10.1063/1.4979016, li.2019.10.1146/annurev-physchem-042018-052416} recently proposed and validated for valence ionization energies (IP-EOM-DSRG).\cite{zhao.2025.10.1021/acs.jctc.5c00992}
A significant challenge in computing core-ionization energies is the fact that X-ray transitions lie deep within the spectrum, while conventional eigenvalue solvers focus mostly on extremal eigenvalues.\cite{coriani.2015.10.1063/1.4935712}
Here we extend IP-EOM-DSRG to core-ionized states using core--valence separation (CVS),\cite{cederbaum.1980.10.1103/PhysRevA.22.206, barth.1981.10.1103/PhysRevA.23.1038} yielding the CVS-IP-EOM-DSRG approach.
The CVS helps target core levels directly, avoiding the costly convergence of many lower-energy valence states and mixing with unphysical states in the continuum.\cite{barth.1985.10.1088/0022-3700/18/5/008, coriani.2015.10.1063/1.4935712, norman.2018.10.1021/acs.chemrev.8b00156, vidal.2019.10.1021/acs.jctc.9b00039}
We assess the CVS-IP-EOM-DSRG approach in combination with both perturbative and non-perturbative approximations to the DSRG transformed Hamiltonian, and introduce a modified MR-DSRG scheme that enforces rigorous core intensivity.

The paper is organized as follows. \Cref{sec:theory} presents the IP-EOM-DSRG framework, the core-intensive extension of the MR-DSRG, and two CVS schemes. \Cref{sec:imp} details the implementation and \Cref{sec:results} reports benchmarks and numerical tests of this new approach. Finally, \Cref{sec:conclusion} summarizes implications and outlook.

\section{Theory}
\label{sec:theory}
\subsection{The EOM-DSRG formalism for photoelectron spectra}
\label{subsec:eom-dsrg}
In this section, we briefly recapitulate the salient features of the IP-EOM-DSRG formalism.
For detailed derivations, we refer the reader to our previous work.\cite{zhao.2025.10.1021/acs.jctc.5c00992}
The IP-EOM-DSRG formalism is an equation-of-motion formulation built on the ground-state multireference DSRG theory.\cite{li.2015.10.1021/acs.jctc.5b00134, li.2016.10.1063/1.4947218, li.2017.10.1063/1.4979016, li.2019.10.1146/annurev-physchem-042018-052416}
In DSRG, the ground state is modeled with a zeroth-order reference state ($\ket{\Phi_0}$) chosen to be a complete- or generalized-active-space (CAS/GAS) wavefunction:
\begin{align}
	\ket{\Phi_0} = \sum_{\mu = 1}^{d}\ket{\phi_\mu} c_{\mu} \label{eq:casscf}
\end{align}
The set of determinants $\mathcal{M} = \{ \ket{\phi_\mu}, \mu =1, \ldots, d \}$ defines the model space, which accounts for the dominant electron configurations relevant to the description of the ground state for a set of molecular geometries, e.g., along a bond-breaking reaction path.
The determinant coefficients $c_{\mu}$ are normalized to one. 
The determinants are formed from a set of spin orbitals $\{\psi_p\}_{p=1}^{N}$  that have been partitioned into core ($\mathbf{C}$, indices $m, n$), active ($\mathbf{A}$, indices $u, v, w, x, y, z$), and virtual ($\mathbf{V}$, indices $e, f$) subsets of size $N_{\mathbf{C}}$, $N_{\mathbf{A}}$, and $N_{\mathbf{V}}$, respectively.
We also introduce two composite orbital subsets: the hole spin orbitals ($\textbf{H} = \textbf{C} \cup \textbf{A}$, indices $i,j,k,l$) and the particle spin orbitals ($\textbf{P} = \textbf{A} \cup \textbf{V}$, indices $a,b,c,d$). 
General spin orbitals (\textbf{G}) are designated as \textit{p}, \textit{q}, \textit{r}, \textit{s}.
We use the notation $\{\hat{a}_{rs\cdots}^{pq\cdots}\} = \{\hat{a}_p^{\dagger}\hat{a}_q^{\dagger} \cdots \hat{a}_s \hat{a}_r\}$ to represent creation ($\hat{a}^{\dagger}$) and annihilation ($\hat{a}$) operator strings normal-ordered with respect to the correlated vacuum $\ket{\Phi_0}$.\cite{mukherjee.1997.10.1016/S0009-26149700714-8, kutzelnigg.1997.10.1063/1.474405, kong.2010.10.1063/1.3439395}

In the MR-DSRG, internally contracted excited configurations are decoupled from the reference $\ket{\Phi_0}$ via a unitary transformation of the Hamiltonian:
\begin{equation}
	\label{eq:dsrg-flow}
	\hat{H} \mapsto \bar{H}(s) = e^{-\hat{A}(s)}\hat{H}e^{\hat{A}(s)}
\end{equation}
where $\bar{H}(s)$ is the MR-DSRG similarity-transformed Hamiltonian, and $\hat{A}(s) = \hat{T}(s) - \hat{T}^{\dagger}(s)$ is an anti-Hermitian combination of the cluster operator $\hat{T}(s)$.
The flow variable $s \in [0, \infty)$ that enters into the definition of $\bar{H}(s)$ plays the role of a regularization parameter responsible for suppressing low-energy excited configurations that lead to numerical issues.\cite{evangelisti.1987.10.1103/PhysRevA.35.4930, kowalski.2000.10.1103/PhysRevA.61.052506, kowalski.2000.10.1002/1097-461X200080:4/5<757::AID-QUA25>3.0.CO;2-A}
Values of $s > 0$ suppress those excitations with energy denominators larger than the energy cutoff $\Lambda = s^{-1/2}$.
The amplitudes are determined by the regularized many-body condition:
\begin{equation}
	\label{eq:manybody-cond}
	[\bar{H}(s)]^{\mathrm{N}} = [e^{-\hat{A}(s)}\hat{H}e^{\hat{A}(s)}]^{\mathrm{N}} = \hat{R}(s)
\end{equation}
where the superscript ``N'' indicates the non-diagonal (excitation rank changing) components of $\bar{H}$ we seek to remove, and $\hat{R}(s)$ is a regularizer that smoothly drives the original Hamiltonian to the one with no coupling between the reference and its excited configurations, \textit{i.e.}, $\lim_{s\rightarrow\infty}[\bar{H}(s)]^{\mathrm{N}}=0$, when all energy denominators are nonzero.
For clarity, in the following text, we omit the symbol ``(s)'' from all $s$-dependent quantities.

The EOM-DSRG approach defines the $\alpha$-th excited state $\ket{\Psi_{\alpha}}$ as:\cite{zhao.2025.10.1021/acs.jctc.5c00992}
\begin{equation}
\ket{\Psi_{\alpha}}=\bar{\mathcal{R}}_{\alpha}\ket{\Psi_0}
\end{equation}
where $\bar{\mathcal{R}}_{\alpha}$ is a state-transfer operator that delivers the $\alpha$-th excited state wavefunction from the ground-state wavefunction $\ket{\Psi_0} = e^{\hat{A}(s)}\ket{\Phi_0}$.
Excited states can be computed by variational minimization of an energy functional augmented with orthonormality constraints:
\begin{equation}
	\label{eq:eom-variational}
\mathcal{L} = \sum_\alpha^{N} \braket{\Psi_0|\bar{\mathcal{R}}_{\alpha}\adj\hat{H}\bar{\mathcal{R}}_{\alpha}|\Psi_0} - \sum_{\alpha\beta}^{N} \lambda_{\alpha\beta} (\braket{\Psi_0|\bar{\mathcal{R}}\adj_{\beta}\bar{\mathcal{R}}_{\alpha}|\Psi_0} - \delta_{\alpha \beta})
\end{equation}
Due to its formal advantages, we employ self-consistent excitation operators introduced by Mukherjee and coworkers,\cite{prasad.1985.10.1103/PhysRevA.31.1287, datta.1993.10.1103/PhysRevA.47.3632} which express $\bar{\mathcal{R}}_{\alpha}$ as a similarity-transformation of a bare excitation operator $\hat{\mathcal{R}}_{\alpha}$:
\begin{equation}
\label{eq:rbar}
	\bar{\mathcal{R}}_{\alpha} \equiv e^{\hat{A}}\hat{\mathcal{R}}_{\alpha}e^{-\hat{A}}
\end{equation}
The bare excitation operator is expanded over a set of excitation operators $\hat{\rho}_p$, with corresponding amplitudes $r^p_{\alpha}$:
\begin{equation}
\label{eq:r operator}
	\hat{\mathcal{R}}_{\alpha} = \sum_{p = 1}^{n_{\mathrm{eom}}}r_{\alpha}^{p}\hat{\rho}_p
\end{equation}
With this choice of $\bar{\mathcal{R}}_{\alpha}$, the energy functional can be expressed in terms of the similarity-transformed Hamiltonian $\bar{H}$ as:
\begin{equation}
\label{eq:eom-master}
\mathcal{L} = \sum_\alpha^{N} \braket{\Phi_0|\hat{\mathcal{R}}_{\alpha}\bar{H}\hat{\mathcal{R}}_{\alpha}|\Phi_0} - \sum_{\alpha\beta}^{N} \lambda_{\alpha\beta} (\braket{\Phi_0|\hat{\mathcal{R}}\adj_{\beta}\hat{\mathcal{R}}_{\alpha}|\Phi_0} - \delta_{\alpha \beta})
\end{equation}
By requiring that all partial derivatives with respect to the excitation amplitudes $r_{\alpha}^p$ (assumed to be real) and the Lagrange multipliers $\lambda_{\alpha\beta}$ to be zero, we arrive at the following generalized eigenvalue problem:
\begin{equation}
	\sum_{q=1}^{n_{\mathrm{eom}}}\braket{\Phi_0|\hat{\rho}_p^{\dagger}\bar{H}\hat{\rho}_q|\Phi_0} r_{\alpha}^q
	= E_{\alpha} \sum_{q=1}^{n_{\mathrm{eom}}} \braket{\Phi_0|\hat{\rho}_p^{\dagger}\hat{\rho}_q|\Phi_0} r_{\alpha}^q
\label{eq:eom_tht}
\end{equation}
where $E_{\alpha}$ is the excited-state energy.
The excitation energy is then given by $\omega_\alpha = E_\alpha - E_0$, where $E_0=\braket{\Phi_0|\bar{H}|\Phi_0}$ is the ground-state energy.

In this work, we truncate the IP-EOM-DSRG excitation operator to one-hole (1h) and two-hole-one-particle (2h1p) operators:
\begin{equation}
	\hat{\mathcal{R}}_{\alpha} = \sum_{i}^{\mathbf{H}}r^i\{\hat{a}_i\}+\frac{1}{2}\sum_{ij}^{\mathbf{H}}\sum_{a}^{\mathbf{P}}r^{ij}_a\{\hat{a}^{a}_{ij}\}
	\label{eq:eom operator}
\end{equation}
The IP-EOM-DSRG formalism is compatible with any choice of underlying MR-DSRG method.
In this work, we investigate the performance of IP-EOM-DSRG-PT2, IP-EOM-DSRG-PT3, and IP-EOM-LDSRG(2), which are based on second- and third-order perturbative MR-DSRG methods (DSRG-MRPT2/3)\cite{li.2015.10.1021/acs.jctc.5b00134, li.2017.10.1063/1.4979016} and an iterative MR-LDSRG(2) formalism.\cite{li.2016.10.1063/1.4947218}

\subsection{Ensuring rigorous core intensivity in IP-EOM-DSRG}
\label{subsec:intensivity_theory}

A basic formal requirement for all excited-state theories (size intensivity) is that excitation energies for localized excitations of a fragment $A$ are unchanged in the presence of other fragments that do not interact with $A$.
It is common to distinguish between full size intensivity and core size intensivity.
The former requires the invariance of excitation energies when the additional non-interacting fragments increase the number of core, active, and virtual orbitals, while the latter only requires invariance with respect to the addition of non-interacting core and virtual orbitals, for a fixed set of active orbitals.

The conditions that guarantee full size intensivity of EOM excitation energies for multireference unitary coupled cluster theory have been discussed in ref~\citenum{li.2025.10.1063/5.0261000}.
Since DSRG relies on the same transformation used in unitary coupled cluster theory, these formal results also find application in the case of EOM-DSRG.
In particular, following Appendix B of ref~\citenum{li.2025.10.1063/5.0261000}, the necessary condition for full size intensivity of IP-EOM-DSRG with up to 2h1p excitations is that matrix elements of $\bar{H}$ between singly excited configurations ($\{\hat{a}_i^a\}\ket{\Phi_0}$) and the reference (the projective residuals $\mathcal{S}_a^i$) are equal to zero, 
 \textit{i.e.}, 
\begin{equation}
\label{eq:singles-projection}
	\mathcal{S}_a^i=\braket{\Phi_0|\{ \hat{a}_a^i \}\bar{H}|\Phi_0} = 0 \quad \forall i \in \mathbf{H} \text{ and } a \in \mathbf{P}
\end{equation}
%NOTE: if the reference is relaxed, then the active-active block of S is automatically zero.

We can evaluate the residuals $\mathcal{S}_a^i$ using Wick's theorem to obtain:\cite{li.2019.10.1146/annurev-physchem-042018-052416}
\begin{equation}
\label{eq:cumulant expansion}
\begin{split}
	\mathcal{S}_a^i = & \sum_j^{\mathbf{H}}\sum_b^{\mathbf{P}} \gamma^{i}_{j} \eta^{b}_{a} \bar{H}_{b}^{j}
 - \frac{1}{2} \sum_j^{\mathbf{H}}\sum_{uxy}^{\mathbf{A}}\gamma^{i}_{j} \lambda^{xy}_{au} \bar{H}_{xy}^{ju} + \cdots \\
  & - \frac{1}{6} \sum_j^{\mathbf{H}}\sum_{uvxyz}^{\mathbf{A}}\gamma^{i}_{j} \lambda^{xyz}_{auv} \bar{H}_{xyz}^{juv} + \cdots
\end{split} 
\end{equation}
All terms in \cref{eq:cumulant expansion} are multiplied by nondiagonal components of $\bar{H}$, which are null for untruncated MR-DSRG theory in the limit of $\lim_{s \rightarrow \infty} [\bar{H}(s)]^\mathrm{N} = 0$.
Therefore, in this case $\mathcal{S}_a^i = 0$ and size intensivity is guaranteed.
For truncated theories at finite $s$ values, the operator $[\bar{H}(s)]^\mathrm{N}$ is not null, and consequently, the singles projective conditions $\mathcal{S}_a^i = 0$ are not satisfied, implying that size intensivity is violated.
As has been shown in ref~\citenum{zhao.2025.10.1021/acs.jctc.5c00992}, full size intensivity errors for IP-EOM-DSRG are small for valence ionization energies of realistic systems.

A possible solution to the size intensivity problem is modifying the ground-state formalism by directly enforcing the projective condition for singles ($\mathcal{S}_a^i=0$), with the remaining double excitation amplitudes solved using the regularized many-body approach (\cref{eq:manybody-cond}).
This approach would be similar to the pIC-MRCC formalism,\cite{datta.2011.10.1063/1.3592494} except for retaining Hermiticity of the transformed Hamiltonian and the presence of a source operator $\hat{R}$ for double excitations (see \cref{eq:manybody-cond}).
One potential disadvantage is the potential for reintroducing numerical instabilities due to small energy denominators.

We propose an alternative approach that aims to rigorously satisfy \textit{core intensivity} of the ionization energies while retaining the numerical stability of the parent MR-DSRG theory.
In most applications, it is paramount that this property is satisfied to guarantee that the ionization energies remain constant as the system of interest is studied in a variety of larger, weakly interacting environments that contribute only additional core and virtual orbitals.
Core intensivity can be satisfied by imposing \cref{eq:singles-projection} only onto the subset of equations that involve core and virtual orbitals, namely, $\mathcal{S}_e^m=0$, $\forall m \in \mathbf{C}$ and $e\in\mathbf{V}$.
The structure of the core-virtual block $\mathcal{S}_e^m$ is also simple, since only the first term on the r.h.s. of \cref{eq:cumulant expansion} survives. The remaining terms are zero because they contain contributions from blocks of $\bar{H}$ that are contracted with null density cumulants (since $m,e \notin \mathbf{A}$, at least one index of the cumulants does not belong to the active orbital set).
Consequently, the condition $\mathcal{S}_e^m=0$ is satisfied if the one-body components $\bar{H}_{e}^{m}$ are null.
To rigorously restore core intensivity in IP-EOM-DSRG, we solve for the condition $\bar{H}_{e}^{m}=0$ for core-virtual singles and impose the MR-DSRG equations (\cref{eq:manybody-cond}) using a finite $s$ value for all other blocks.
Note that one may equivalently view this approach as taking the $s \rightarrow \infty$ limit of the singles core-virtual block of the original MR-DSRG equations.
This approach is expected to be intruder-free, as long as core and virtual orbitals are energetically well-separated, which is typically ensured by appropriate active space selection.
In addition, this approach is anticipated to introduce only negligible changes to the ground-state energy.
This aspect will be investigated numerically in \Cref{subsec:intensivity}.

\subsection{IP-EOM-DSRG with core-valence separation for X-ray photoelectron spectra}
\label{subsec:cvs}

\begin{figure*}[!htb]
\centering
\includegraphics[width=6.25in]{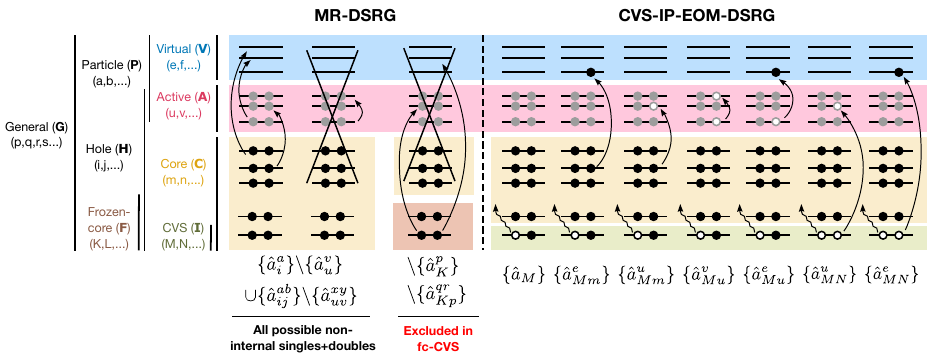} 
\caption{Orbital spaces and schematic definitions of the cluster operator ($\hat{T}$, left) and the EOM excitation operator ($\hat{\mathcal{R}}$, right) used in the MR-DSRG and CVS-IP-EOM-DSRG schemes. Curved arrows represent electron excitations, while wavy arrows denote electron ionizations. Crossed-out terms are excluded from the definition of an operator.
\label{fig:cvs}
}
\end{figure*}

IP-EOM-DSRG theories can simulate electron ionization involving non-active molecular orbitals, including core-ionized states.
However, core-ionized states are numerically challenging to access as they are deeply embedded in the autodetaching continuum.
To address this issue, we adopt the core-valence separation (CVS), where the continuum is projected out, and the core-ionized states are stabilized.
We refer to this combination of methods as CVS-IP-EOM-DSRG.
First proposed by Cederbaum et al.,\cite{cederbaum.1980.10.1103/PhysRevA.22.206} the CVS approximation was later used for simulating core-level excitation and ionization with a variety of electronic structure theories, including coupled-cluster theory,\cite{coriani.2015.10.1063/1.4935712, vidal.2019.10.1021/acs.jctc.9b00039, liu.2019.10.1021/acs.jctc.8b01160, vidal.2020.10.1039/C9CP03695D, vidal.2020.10.1021/acs.jpclett.0c02027, thielen.2021.10.1063/5.0047134} single reference and multireference ADC theory,\cite{schirmer.2001.10.1063/1.1418437, thiel.2003.10.1063/1.1584658, wenzel.2014.10.1002/jcc.23703a, wenzel.2015.10.1063/1.4921841, demoura.2022.10.1039/D1CP05476G, mazin.2023.10.1021/acs.jctc.3c00477, demoura.2024.10.1021/acs.jpca.4c03161} second-order excited-state perturbation theory,\cite{garner.2020.10.1063/5.0020595} linear-response CASSCF,\cite{helmich-paris.2021.10.1002/qua.26559} and linear-response density cumulant theory.\cite{peng.2019.10.1021/acs.jpca.8b12259}

In this work, we test two different approaches to CVS that differ in the treatment of core orbitals in the ground-state MR-DSRG step.
We distinguish 1) an edge-specific CVS set $\mathbf{I}$ (indices $M, N$) used to restrict the EOM excitation operator, and 2) a frozen-core set $\mathbf{F}$ used in the ground-state MR-DSRG calculation, with $\mathbf{I}\subseteq \mathbf{F}$ in practice.
In the full-CVS scheme, all electrons are correlated in the ground-state MR-DSRG ($\mathbf{F}$ is empty), and the EOM operator is restricted to excitations involving at least one orbital from $\mathbf{I}$.
This is equivalent to the approach of Coriani and Koch.\cite{coriani.2015.10.1063/1.4935712}
In the fc-CVS scheme, the ground-state MR-DSRG computation is performed with a fixed frozen-core set $\mathbf{F}$, which typically includes all 1s orbitals for atoms whose K-edges will be probed.
These orbitals are excluded from the cluster amplitudes, while the EOM operator is still restricted using the smaller set $\mathbf{I}$.
This follows the idea of Cederbaum et al.\cite{cederbaum.1980.10.1103/PhysRevA.22.206} and matches the approach of Vidal et al.\cite{vidal.2019.10.1021/acs.jctc.9b00039}
For example, in \ce{CO}, we choose $\mathbf{F} = \mathrm{\{C\ 1s, O\ 1s\}}$.
For the C K-edge spectrum, $\mathbf{I} = \mathrm{\{C\ 1s\}}$, and for the O K-edge, $\mathbf{I} = \mathrm{\{O\ 1s\}}$.
This ensures that both edges are computed from the same ground-state reference wavefunction, while the EOM step targets only the desired edge.
\Cref{fig:cvs} shows orbital spaces and schematic definitions of the cluster operator and the EOM excitation operator used in the MR-DSRG and CVS-IP-EOM-DSRG schemes.

The intensity of a photoelectron transition to the $\alpha$-th state with excitation energy $\omega_\alpha$ can be approximated using the spectroscopic factor $S_{\alpha}$,\cite{meldner.1971.10.1103/PhysRevA.4.1388, kheifets.1994.10.1088/0953-4075/27/15/005, chatterjee.2019.10.1021/acs.jctc.9b00528} which is given by:
\begin{equation}
\begin{split}
	S_{\alpha}&=\sum_{M}^{\mathbf{CVS}}|\braket{\Psi^{N-1}_{\alpha}|\hat{a}_M|\Psi_0^{N}}|^2 
	=\sum^{\mathbf{CVS}}_M|\braket{\Phi_0|\hat{\mathcal{R}}_{\alpha}\adj \bar{a}_M |\Phi_0}|^2
\end{split}
\end{equation}
where $\bar{a}_M = e^{-\hat{A}}\hat{a}_M e^{\hat{A}}$ is a similarity-transformed operator.
In the fc-CVS scheme, where $\hat{A}$ does not contain excitation operators involving CVS orbitals, $\hat{a}_M$ and $\hat{A}$ commute, and this expression simplifies to:
\begin{equation}
\label{eq:working S}
	S_{\alpha} = \sum^{\mathbf{CVS}}_M|\braket{\Phi_0|\hat{\mathcal{R}}_{\alpha}\adj \hat{a}_M |\Phi_0}|^2
\end{equation}
In the full-CVS scheme, \cref{eq:working S} serves as a zeroth-order approximation to the spectroscopic factor.
In this work, we employ \cref{eq:working S} for both CVS schemes.

Comparing to the IP-EOM-DSRG formalism, major computational savings are achieved by neglecting excitations that include only active orbitals.
In the scenario where $N_{\mathbf{V}} > N_{\mathbf{C}} \gg N_{\mathbf{A}}$, the dominant computational cost of CVS-IP-EOM-DSRG scales as $\mathcal{O}(N_{\mathbf{I}}N_{\mathbf{C}}^2N_{\mathbf{V}}^2)$ (quartic overall in basis size).
This scaling arises from contracting the $\mathbf{CVCV}$ block of the similarity-transformed Hamiltonian $\bar{H}$ with the operator $\hat{a}^{\mathbf{V}}_{\mathbf{IC}}$.
Additionally, the cost associated with the active space scales as $\mathcal{O}(N_{\mathbf{I}}N_{\mathbf{A}}^8)$, arising from the contraction between the $\mathbf{AAAA}$ block of the $\bar{H}$ with the operator $\hat{a}^{\mathbf{A}}_{\mathbf{IA}}$.
For brevity, we will remove the `CVS-IP' prefix in the abbreviations of CVS-IP-EOM-DSRG theories henceforth.

\section{Implementation}
\label{sec:imp}
All EOM-DSRG computations are performed using \textsc{NiuPy},\cite{Li_niupy_Open-source_implementation_2025} an open-source Python library of multireference electronic structure theories for simulating excited states.
The underlying CASSCF and MR-DSRG ground-state computations are carried out with a development version of the \textsc{Forte} quantum chemistry package.\cite{evangelista.2024.10.1063/5.0216512}
\textsc{NiuPy} serves as a general-purpose MR-EOM solver that is agnostic to the ground-state method, requiring only the effective Hamiltonian and reduced density matrices to solve \cref{eq:eom_tht}.
To speed up both the derivation and implementation of EOM-DSRG, we use a development version of the \textsc{Wick\&d} software package,\cite{evangelista.2022.10.1063/5.0097858} which generates working equations and executable code at runtime.

In our implementation, the generalized eigenvalue problem in \cref{eq:eom_tht} is transformed into a standard eigenvalue problem, $\mathbf{\tilde{H}\tilde{r} = E\tilde{r}}$, where $\mathbf{\tilde{H}} = \mathbf{S}^{-1/2}\mathbf{\bar{H}}\mathbf{S}^{-1/2}$ and $\mathbf{\tilde{r}} = \mathbf{S}^{1/2}\mathbf{r}$.
The matrix square root $\mathbf{S}^{1/2}$ is computed efficiently by exploiting the block-diagonal structure of $\mathbf{S}$.
For each diagonal block, we apply a threshold $\eta_{\mathrm{d}} = 10^{-10}$ to eliminate linear dependencies, except for the block corresponding to \textit{spectator} excitations,\cite{jeziorski.1981.10.1103/PhysRevA.24.1668} where operators $\hat{a}_M$ and $\hat{a}_{Mu}^v$ are known to exhibit strong linear dependence.
For this block, we adopt the sequential orthogonalization procedure developed by Hanauer and K{\"o}hn,\cite{hanauer.2011.10.1063/1.3592786} where $\hat{a}_M$ operators are first orthogonalized and then projected out from the $\hat{a}_{Mu}^v$ operator space.
Subsequently, $\hat{a}_{Mu}^v$ operators are orthogonalized using a larger threshold of $\eta_{\mathrm{s}} = 10^{-5}$.
This orthogonalization strategy has also been used in the implementation of MR-ADC.\cite{chatterjee.2019.10.1021/acs.jctc.9b00528, chatterjee.2020.10.1021/acs.jctc.0c00778}
The resulting eigenvalue problem is solved using a multiroot implementation of the Davidson algorithm,\cite{davidson.1975.10.1016/0021-99917590065-0, liu1978simultaneous} which avoids the explicit storage of $\mathbf{\tilde{H}}$.

\section{Results and discussion}
\label{sec:results}
\subsection{CVS-IP scheme and flow parameter choice}
\label{subsec:cvs and flow}
To calibrate the EOM-DSRG methods, we first examine the sensitivity of the vertical K-edge ionization energies with respect to the choice of the CVS scheme.
For this purpose, in \Cref{tab:cvs} we report eight ionization energies for \ce{HF}, \ce{CO}, \ce{N2}, \ce{F2}, and \ce{H2O} computed using different levels of EOM-DSRG theory and the two CVS schemes.
Results are benchmarked against highly-accurate CVS-EOMIP-CCSDTQ reference data.\cite{liu.2019.10.1021/acs.jctc.8b01160}
All EOM-DSRG computations use $s = 0.5\ E_{\mathrm{h}}^{-2}$ and the core-intensive modification of the MR-DSRG formalism.
To enable a direct comparison, our EOM-DSRG computations adopt the same geometries, basis sets, and exact two-component relativistic treatment as in the CVS-EOMIP-CCSDTQ study.
CVS-EOMIP-CCSDTQ calculations were carried out using the full-CVS scheme.
For each method, we summarize the error statistics by computing the mean absolute error (MAE) and standard deviation (STD).

As shown in \Cref{tab:cvs}, computations using the full-CVS scheme consistently overestimate the ionization energy for all truncation levels by more than 1~eV.
Within the full-CVS scheme, the DSRG-MRPT2 level of theory yields the smallest MAE (1.01~eV), whereas MRPT3 and MR-LDSRG(2) give larger MAEs of 2.02 and 1.52~eV, respectively.
This behavior is consistent with error cancellation: an incomplete treatment of dynamical electron correlation in DSRG-MRPT2 in the ground state likely offsets the missing explicit orbital relaxation in the EOM step.
Higher-level ground-state treatments reduce that cancellation and expose the intrinsic bias of full CVS (to inflate excitation energies), which suppresses core-valence relaxation/screening. 
In contrast, the standard deviation is comparable across the methods, ranging from 0.73~eV for EOM-DSRG-PT3 to 0.67~eV for EOM-LDSRG(2).

Imposing the frozen-core approximation systematically reduces the ionization energy, leading to improved agreement with the CVS-EOMIP-CCSDTQ reference across all levels of theory, consistent with the error cancellation pointed out by Vidal et al.\cite{vidal.2019.10.1021/acs.jctc.9b00039}
Under the fc-CVS scheme, the highest-level EOM-LDSRG(2) method yields the smallest MAE of 0.58~eV, while EOM-DSRG-PT2 and EOM-DSRG-PT3 yield MAEs of 0.71 and 0.73~eV, respectively.
Notably, within the fc-CVS scheme, EOM-DSRG-PT2 and EOM-DSRG-PT3 exhibit large errors for \ce{F2}, a notoriously challenging diatomic;\cite{bytautas.2007.10.1063/1.2800017} however, these errors are reduced by EOM-LDSRG(2).
This trend is consistent with our earlier observations,\cite{zhao.2025.10.1021/acs.jctc.5c00992} indicating that a higher-level treatment of dynamic correlation is needed for this system.
Due to the lower statistical errors of the fc-CVS for this benchmark set, all subsequent EOM-DSRG computations employ this scheme.

Another aspect we examine is the sensitivity of the ionization energy and spectroscopic factor to the DSRG flow parameter $s$.
As an example, in \Cref{tab:flow parameter} we report the carbon K-edge ionization energy and spectroscopic factor of \ce{CO}, computed using various levels of EOM-DSRG theory with $s$ values ranging from 0.25 to 4 $E_{\mathrm{h}}^{-2}$, employing the cc-pCVTZ-DK basis set and the 1e-sf-X2C relativistic correction.
Across all levels of theory, the ionization energy and spectroscopic factor exhibit weak dependence on $s$ until it reaches a relatively large value ($s = 2.0\ E_{\mathrm{h}}^{-2}$), beyond which the MR-LDSRG(2) ground-state amplitude equations fail to converge. 
An analysis of the double substitution amplitudes involving both $\alpha$ and $\beta$ electrons at large $s$ values shows amplitudes as large as 0.6, indicating the presence of intruder states,\cite{zhang.2019.10.1021/acs.jctc.9b00353} while for $s \leq 1.0\ E_{\mathrm{h}}^{-2}$ all amplitudes are less than 0.1.
Within the recommended range of $s \in [0.5, 1.0]\ E_{\mathrm{h}}^{-2}$, the EOM-DSRG ionization energy remains nearly constant at the MR-LDSRG(2) level, varying by only 0.02~eV.
At the DSRG-MRPT2 and DSRG-MRPT3 levels, the ionization energy shows weak dependence, with maximum variations of 0.09 and 0.08~eV, respectively.
Based on these findings and prior studies that suggest an optimal $s$ value in the range [0.5,1.0]~$E_{\mathrm{h}}^{-2}$,\cite{li.2016.10.1063/1.4947218, li.2017.10.1063/1.4979016, huang.2024.10.1021/acs.jctc.4c00835, li.2019.10.1146/annurev-physchem-042018-052416} we adopt the value $s = 0.5\ E_{\mathrm{h}}^{-2}$ for all subsequent EOM-DSRG computations.

\begin{table*}[ht!]
	\caption{K-edge ionization energy errors (in eV) for EOM-DSRG methods [PT2 = DSRG-MRPT2, PT3 = DSRG-MRPT3, and LDSRG(2) = MR-LDSRG(2)] computed using two CVS schemes: full-CVS (full) and fc-CVS (fc). All results are reported relative to the CVS-EOMIP-CCSDTQ (CCSDTQ) reference values taken from ref~\citenum{liu.2019.10.1021/acs.jctc.8b01160}. The element symbols in bold indicate the atom from which the core electron is ionized.
For homonuclear diatomics, we distinguish ionization from symmetric ($g$) and antisymmetric ($u$) linear combinations of the atomic 1s orbitals.
	}
	\label{tab:cvs}
	\setlength{\extrarowheight}{2pt}
	\setstretch{1}
	\centering
	\begin{threeparttable}
	\begin{tabular}{lccccccc}
		\hline
		
		\hline
		 & CCSDTQ &  \multicolumn{2}{c}{PT2} & \multicolumn{2}{c}{PT3} & \multicolumn{2}{c}{LDSRG(2)}\\ \cmidrule(lr){3-4} \cmidrule(lr){5-6} \cmidrule(lr){7-8} 
		& & fc & full & fc & full & fc & full \\ \hline
		\ce{H\textbf{F}} & 694.45 & -1.21 & 0.19 & 0.41 & 1.86 & -0.96 & 0.39\\
		\ce{\textbf{C}O} & 296.30 & 0.14 & 1.93 & -0.14 & 1.22 & -0.16 & 1.21\\
		\ce{C\textbf{O}} & 542.68 & -1.24 & 1.93 &-0.09 & 1.29 & -0.45 & 0.91\\
		\ce{\textbf{N}2} ($^2\Sigma_g^+$) & 410.03 & 0.05 & 1.38 & 0.43 & 1.72 & 0.32 & 1.61\\
		\ce{\textbf{N}2} ($^2\Sigma_u^+$) & 409.95 & 0.03 & 1.37 & 0.41 & 1.70 & 0.30 & 1.59\\
		\ce{\textbf{F}2} ($^2\Sigma_g^+$) & 696.82 & -1.00 & 0.42 & 1.73 & 3.08 & 0.98 & 2.31\\
		\ce{\textbf{F}2} ($^2\Sigma_u^+$) & 696.81 & -1.01 & 0.43 & 1.72 & 3.08 & 0.97 & 2.31\\
		\ce{H2\textbf{O}} & 539.99 & -0.97 & 0.42 & 0.90 & 2.23 & 0.52 & 1.84\\
		MAE & -- & 0.71 & 1.01 & 0.73 & 2.02 & 0.58 & 1.52\\
		STD & -- & 0.61 & 0.72 & 0.73 & 0.73 & 0.68 & 0.67\\
		\hline	
			
		\hline
	\end{tabular}
	\end{threeparttable}
\end{table*}

\begin{table}[ht!]
	\caption{Carbon K-edge core-ionization energy ($\omega$, in eV) and spectroscopic factor ($S$) of \ce{CO} computed with various levels of EOM-DSRG theory [PT2 = DSRG-MRPT2, PT3 = DSRG-MRPT3, and LDSRG(2) = MR-LDSRG(2)] using different flow parameter values ($s$, in $E_{\mathrm{h}}^{-2}$).}
	\label{tab:flow parameter}
	\setlength{\extrarowheight}{2pt}
	\setstretch{1}
	\centering
	\begin{threeparttable}
	\begin{tabular}{lcccccc}
		\hline
		
		\hline
		$s$ & \multicolumn{2}{c}{PT2} & \multicolumn{2}{c}{PT3} & \multicolumn{2}{c}{LDSRG(2)}\\ \cmidrule(lr){2-3} \cmidrule(lr){4-5} \cmidrule(lr){6-7} 
		& $\omega$& $S$ & $\omega$& $S$ & $\omega$& $S$ \\ \hline
		0.25 & 296.08 & 1.5477 & 296.02 & 1.5522 & 296.04 & 1.5564\\
		0.50 & 296.11 & 1.5421 & 296.16 & 1.5522 & 296.14 & 1.5547\\
		1.00 & 296.02 & 1.5348 & 296.24 & 1.5574 & 296.12 & 1.5531\\
		2.00 & 295.89 & 1.5256 & 296.27 & 1.5607 & --$^a$ & --$^a$\\
		4.00 & 295.62 & 1.5097 & 296.27 & 1.5496 & --$^a$ & --$^a$\\
		\hline	
		
		\hline
	\end{tabular}
	\end{threeparttable} \\
	$^a$ MR-LDSRG(2) computation on the ground state fails to converge.
\end{table}

\subsection{Size intensivity}
\label{subsec:intensivity}
In this section, we carefully examine the core intensivity and quantify the size-intensivity error of EOM-DSRG.
As discussed in \Cref{subsec:intensivity_theory}, although the EOM-DSRG formalism is not size-intensive, rigorous core intensivity can be achieved by solving a set of modified MR-DSRG equations.
To begin, we discuss the impact of the core-intensive modification of the MR-DSRG on the ground-state energy.
For this purpose, we consider the \ce{CO} molecule along the dissociation path.
To properly describe the atomic asymptotic limit, we include all 2p orbitals of C and O in the active space while including all 2s orbitals in the core space. 
This setup is designed to artificially increase the magnitude of one-body core-virtual components $\bar{H}_{e}^{m}$.

In  \Cref{tab:separate_s}, we compare results from the original MR-DSRG framework with those from the modified MR-DSRG equations where we set $\bar{H}_{e}^m = 0$.
As shown in \Cref{tab:separate_s}, the core-intensive formulation of MR-DSRG has a negligible effect on the MR-DSRG ground-state energy. 
The largest deviation, observed in the MR-LDSRG(2) calculation at the 4.0 \AA\ geometry, is approximately $7.0 \times 10^{-5} E_{\mathrm{h}}$. 
The impact on the core-ionization energy is also small, with the largest deviation being about 0.037~eV, occurring at the 1.0 \AA\ geometry in the MR-LDSRG(2) calculation. 

We then test core intensivity numerically. 
We compute the vertical core and valence ionization energies of \ce{HF} at the equilibrium ($r_\mathrm{e}$) and stretched (2$r_\mathrm{e}$) geometries using EOM-DSRG-PT3, in the presence of an increasing number of non-interacting helium atoms.
We use the cc-pCVTZ-DK basis and a scalar 1e-sf-X2C relativistic treatment, the same computational strategy that will be used for benchmarking our methods in \Cref{subsec:benchmark}.
The active space contains 6 electrons and 5 active orbitals (H 1s, F 2s/2p orbitals).
In all our numerical tests, the core-intensivity error is within the convergence threshold used to converge the EOM-DSRG excitation energies ($< 10^{-9}$~eV), confirming that the EOM-DSRG formalism is core-intensive.

We also examine the full-intensivity error by computing the vertical core-ionization energies of the \ce{HF} + \ce{HF} composite system.
The active space for the composite system contains 12 electrons and 10 orbitals, double the size of the single \ce{HF} subsystem.
The full-intensivity error of EOM-DSRG-PT3 is 15.7~meV at the equilibrium geometry and 6.3~meV at the stretched geometry, both at least one order of magnitude smaller than the intrinsic error of the method, as will be shown in \Cref{subsec:benchmark}.

\begin{table*}[!htb]
	\caption{Ground-state MR-DSRG energy ($E_{\mathrm{DSRG}}$, in $E_{\mathrm{h}}$) and core-ionization energy (CVS-IP, in eV) for \ce{CO} at each geometry computed using EOM-DSRG methods [PT2 = DSRG-MRPT2, PT3 = DSRG-MRPT3, and LDSRG(2) = MR-LDSRG(2)].
	Results for the original theory (``Original'') and a modified ground-state MR-DSRG approach that solves for the condition $\bar{H}_{e}^{m} = 0$ and ensures core intensivity of the ionization energies (``Core-intensive''). All results use a value of $s = 0.5~E_{\mathrm{h}}^{-2}$.}
	\label{tab:separate_s}
	\setlength{\extrarowheight}{2pt}
	\setstretch{1.2}
	\centering
	\small
	\begin{threeparttable}
	\begin{tabular}{ccccccc}
		\hline
		
		\hline
		\multirow{2}{*}{Geometry (\AA)} & \multicolumn{2}{c}{PT2} & \multicolumn{2}{c}{PT3} & \multicolumn{2}{c}{LDSRG(2)}\\  \cmidrule(lr){2-3} \cmidrule(lr){4-5} \cmidrule(lr){6-7} 
		& Original & Core-intensive & Original & Core-intensive & Original & Core-intensive \\ \hline
		\multicolumn{7}{c}{$E_{\mathrm{DSRG}}$ ($E_{\mathrm{h}}$) } \\
		1.0 & -113.136378 & -113.136378 & -113.157834 & -113.157836 & -113.169934 & -113.170001\\
		2.0 & -112.893327 & -112.893329 & -112.901037 & -112.901056 & -112.907672 & -112.907739 \\
		3.0 & -112.840977 & -112.840977 & -112.828473 & -112.828499 & -112.832786 & -112.832832\\
		4.0 & -112.801600 & -112.801600 & -112.810382 & -112.810382 & -112.816276 & -112.816346 \\
		5.0 & -112.807485 & -112.807485 & -112.814293 & -112.814293 & -112.817197 & -112.817201 \\
		\multicolumn{7}{c}{CVS-IP (eV)} \\		
		1.0 & 295.937 & 295.963 & 295.629 & 295.618 & 295.675 & 295.637 \\
		2.0 & 298.864 & 298.878 & 298.597 & 298.585 & 298.522 & 298.492 \\
		3.0 & 298.571 & 298.598 & 297.951 & 297.945 & 298.061 & 298.028 \\
		4.0 & 297.560 & 297.588 & 297.368 & 297.362 & 297.340 & 297.305 \\
		5.0 & 297.390 & 297.390 & 297.238 & 297.238 & 297.235 & 297.235 \\
		\hline	
		
		\hline
	\end{tabular}
	\end{threeparttable}
\end{table*}

\subsection{Vertical ionization energies}
\label{subsec:benchmark}
In this section, we benchmark the accuracy of the EOM-DSRG methods using a set of sixteen medium-sized molecules previously studied by Liu et al.\cite{liu.2019.10.1021/acs.jctc.8b01160}
We use the same geometries, with diatomic molecules taken from experiment and polyatomic molecules optimized at the SFX2C-1e-CCSD(T)/cc-pCVQZ level of theory.
We benchmark the EOM-DSRG methods against various levels of state-specific and state-averaged GAS-DSRG theory, as well as SR-ADC, MR-ADC, and CVS-EOM-CCSDT.
All EOM-DSRG computations employ the cc-pCVTZ-DK basis set and a scalar 1e-sf-X2C relativistic treatment, consistent with the GAS-DSRG study.\cite{huang.2024.10.1021/acs.jctc.4c00835}
Reference data for SR-ADC and MR-ADC are taken from the work by De Moura and Sokolov,\cite{demoura.2022.10.1039/D1CP05476G} while CVS-EOM-CCSDT results are from Liu et al.\cite{liu.2019.10.1021/acs.jctc.8b01160}
In those studies, the cc-pCVTZ basis set was used for nonrelativistic computations, and a recontracted 1e-sf-X2C basis set was employed when scalar relativistic effects were included.
The detailed choice of active spaces and the raw data used in this section are listed in the Supporting Information.
We assess the accuracy of each method by calculating errors with respect to experimental values.
We note that the experimental ionization energies used for comparison are taken from studies that lack the resolution to resolve vibrational levels, and therefore approximately correspond to vertical ionization energies.

\begin{figure*}[!htb]
\centering
\includegraphics[width=6.25in]{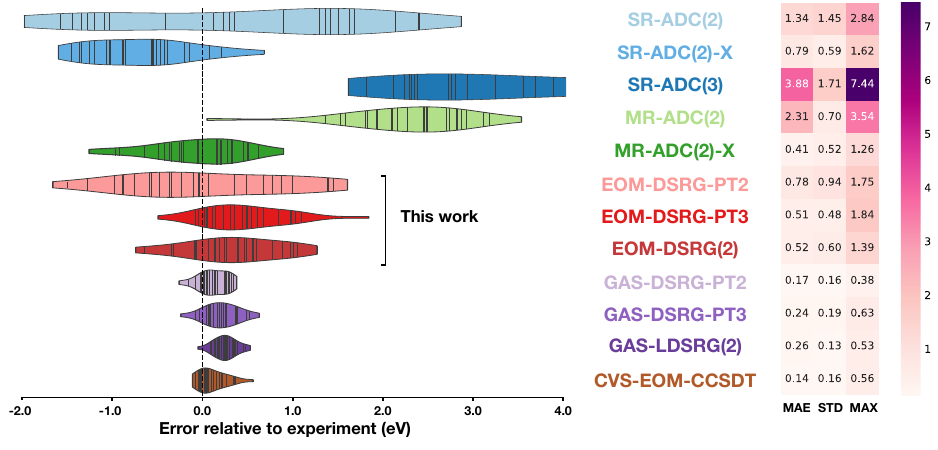} 
\caption{Violin plots of the errors in vertical core-ionization energies for a test set of 16 molecules, computed using various electronic structure methods and referenced against experimental values.
\label{fig:error}
}
\end{figure*}

In \Cref{fig:error}, we present the error distributions and statistics for a range of methods, including SR-ADC, MR-ADC, EOM-DSRG, GAS-DSRG, and CVS-EOM-CCSDT.
All EOM-DSRG methods show good agreement with experiment, with MAEs below 0.8~eV and STDs below 0.9~eV.
Among them, EOM-DSRG-PT3 and EOM-LDSRG(2) exhibit nearly identical accuracy, with MAEs of 0.52 and 0.51~eV and STDs of 0.49 and 0.52~eV, respectively.
Compared to MR-ADC methods,\cite{demoura.2022.10.1039/D1CP05476G} which are based on a similar multideterminantal many-body expansion, EOM-DSRG approaches significantly outperform the strict second-order MR-ADC(2) (MAE = 2.31~eV, STD = 0.70~eV) and offer accuracy comparable to the extended MR-ADC(2)-X (MAE = 0.41~eV, STD = 0.52~eV).
The maximum absolute error (MAX) of each EOM-DSRG method is also comparable to that of MR-ADC(2)-X (all below 2~eV), and significantly smaller than that of MR-ADC(2), which reaches 3.54~eV.
As expected, EOM-DSRG-PT2 yields larger errors (MAE = 0.74~eV, STD = 0.90~eV) compared to its higher-level counterparts.

When compared to SR-ADC methods, EOM-DSRG theories are generally more accurate. 
Notably, the performance of single-reference methods improves substantially upon inclusion of 3h2p-type excitations, as evidenced by the CVS-EOM-CCSDT method, which achieves a MAE of just 0.14~eV.
However, this comes at the cost of increased computational scaling of $\mathcal{O}(N_{\mathbf{I}}N_{\mathbf{C}}^2N_{\mathbf{V}}^4)$.
These findings suggest that, for molecules at equilibrium geometries, the improvements in core-ionization energies are mostly brought by the higher-level description of the dynamical correlation.
Finally, GAS-DSRG methods are more accurate than SR-ADC, MR-ADC, and EOM-DSRG theories, and their accuracy is similar to that of CVS-EOM-CCSDT.
This improved performance is primarily attributed to the explicit orbital optimization of the core-ionized state.

Overall, our benchmark results demonstrate that EOM-DSRG methods accurately predict K-edge core-ionization energies across the test set, with the performance trend approximately following EOM-LDSRG(2) $\approx$ EOM-DSRG-PT3 $>$ EOM-DSRG-PT2.
In particular, the DSRG-MRPT3 level of theory offers a good balance between accuracy and computational efficiency, making it a reliable choice for modeling core-ionized states.

\subsection{Potential energy curves and vibrationally resolved X-ray photoelectron spectra}
\label{subsec:pes}

After benchmarking the EOM-DSRG methods on vertical core-ionization energies, we focus on applications that require a multireference treatment to accurately model ground and core-ionized states far from the equilibrium geometry.
To this end, we extend our core ionization computations to full potential energy curves of diatomics and simulate the corresponding vibrationally-resolved XPS, comparing theoretical results with experimental data.
We focus on the bond dissociation curves of \ce{N2} and \ce{CO}, comparing our results against MR-ADC(2)-X and state-specific GAS-DSRG.
For \ce{N2}, we use reference data computed at the MR-LDSRG(2) truncation level, while for \ce{CO}, we employ the DSRG-MRPT3 level due to the lack of convergence of the GAS-LDSRG(2) procedure.
Both GAS-DSRG-PT3 and GAS-LDSRG(2) incorporate dynamical correlation effects beyond second-order perturbation theory in a state-specific manner. Previous benchmarks have shown that GAS-DSRG-PT3 and GAS-LDSRG(2) provide highly accurate core-ionization energies across various sizes of molecules, with MAEs of about 0.3 eV compared to experiment, while also yielding reliable potential energy surfaces.\cite{huang.2024.10.1021/acs.jctc.4c00835} GAS-DSRG has also been adopted as a benchmark reference in a previous MR-ADC study.\cite{mazin.2023.10.1021/acs.jctc.3c00477}
MR-ADC(2)-X computations are performed using the \textsc{Prism} software package,\cite{prism} while GAS-DSRG computations are carried out with the \textsc{Forte} software package.\cite{evangelista.2024.10.1063/5.0216512}
All computations employ the cc-pCVTZ-DK basis set and the 1e-sf-X2C relativistic treatment.

\begin{figure*}[!htb]
\centering
\includegraphics[width = 5.25in]{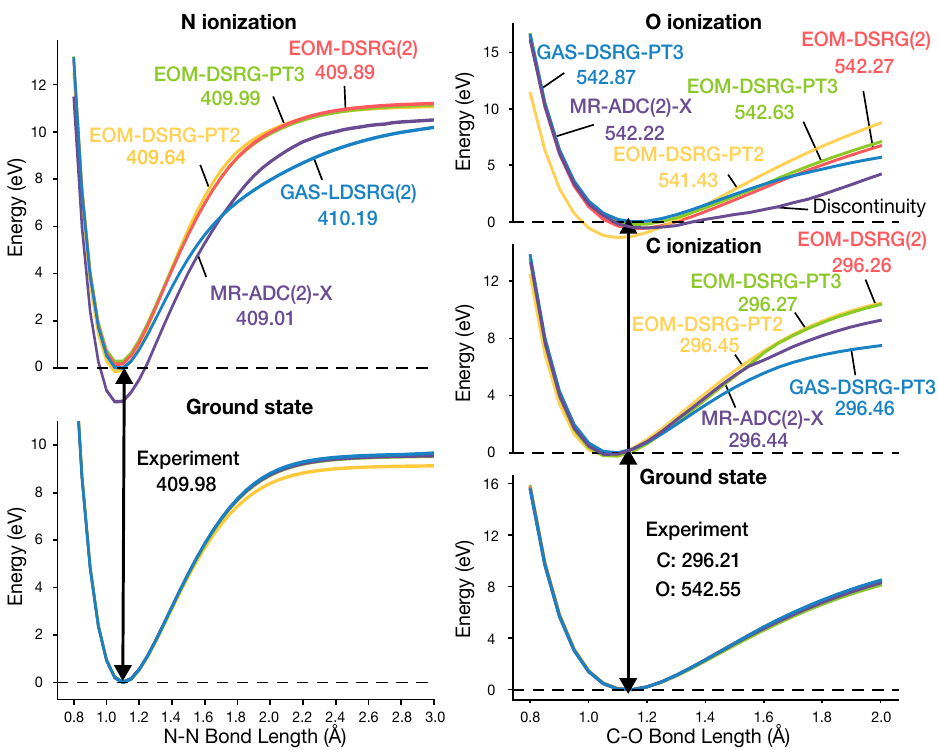} 
\caption{Potential energy curves for the ground and K-edge core-ionized states of \ce{N2} (left) and \ce{CO} (right) computed using various levels of EOM-DSRG, GAS-DSRG, and MR-ADC(2)-X. All energies shown in the plots are vertical ionization energies in units of~eV. For \ce{N2}, the $^2\Sigma_u^+$ core-ionized state is shown; for \ce{CO}, both C 1s and O 1s core-ionized states are included. All curves are shifted such that the energy minima of the GAS-DSRG curves in all panels are set to zero.
Additionally, all methods except GAS-DSRG are further shifted so that their ground-state energy minima align with the minimum of GAS-DSRG.
\label{fig:pes}
}
\end{figure*}

\Cref{fig:pes} shows potential energy curves (PECs) for the ground and core-ionized states of \ce{N2} (N K-edge) and \ce{CO} (C and O K-edge) computed using different methods.
All EOM-DSRG methods follow the reference GAS-LDSRG(2) curve for \ce{N2} at short bond distances [$r(\mathrm{N-N}) \leq 1.3$ {\AA}], with deviations consistently within 0.50~eV, indicating accurate descriptions near equilibrium.
In contrast, MR-ADC(2)-X deviates the most in this region, underestimating the vertical ionization energy by 1.18~eV relative to the GAS-LDSRG(2) result.
As the bond is stretched, all methods continue to yield qualitatively correct PECs.
EOM-DSRG curves remain nearly indistinguishable, whereas the MR-ADC(2)-X curve runs almost parallel but lies consistently lower in energy.
In the strongly correlated dissociation limit, EOM-DSRG methods and MR-ADC(2)-X overestimate the ionization energy.
This is because GAS-LDSRG(2) includes double excitations in the $(N-1)$ electron Hilbert space, effectively capturing 3h2p-type correlation effects that are important for accurately describing core-ionized states.

For the \ce{CO} system, the EOM-DSRG methods show good agreement with the GAS-DSRG-PT3 reference for the C 1s core-ionized state near the equilibrium geometry [$r(\mathrm{C-O}) \leq 1.5$ {\AA}]. 
However, EOM-DSRG-PT2 significantly underestimates the O 1s ionization energy by 1.44~eV.
EOM-DSRG-PT3 and EOM-LDSRG(2) yield nearly identical results across all states and consistently overestimate the ionization energy in the stretched-bond regime.
MR-ADC(2)-X exhibits pronounced discontinuities in both the C and O K-edge core-ionized PECs.
This issue is a known limitation of internally contracted methods that involve amplitude truncation,\cite{li.2025.10.1063/5.0261000} though it is less pronounced in EOM-DSRG methods for the \ce{CO} system.

To further explore the accuracy of the EOM-DSRG PECs, we compute vibrational constants for all computed core-ionized PECs in \Cref{tab:spec}, obtained by fitting the PECs to Morse potentials.
For the \ce{N2} system, all methods reproduce the experimentally observed bond-length contraction upon core ionization with reasonable accuracy.\cite{ueda.2009.10.1140/epjst/e2009-00978-7}
GAS-LDSRG(2) achieves the best agreement, with differences in bond lengths ($\Delta r_e$) deviating by less than 0.001 {\AA} from experiment.
EOM-DSRG-PT2 shows the largest deviations, with $\Delta r_e$ differing from experiment by $-0.0110$ {\AA} and $-0.0127$ {\AA} for the $^2\Sigma_u^+$ and $^2\Sigma_g^+$ states, respectively.
GAS-LDSRG(2) performs well also for vibrational frequencies, deviating by less than 5 cm$^{-1}$ from experiment.
EOM-DSRG-PT2 and MR-ADC(2)-X exhibit larger errors (approximately 160 cm$^{-1}$), while EOM-DSRG-PT3 and EOM-LDSRG(2) reduce the error to below 100 cm$^{-1}$.

For the \ce{CO} system, all methods correctly predict bond-length contraction upon C 1s core ionization.\cite{matsumoto.2006.10.1016/j.cplett.2005.09.094}
EOM-DSRG-PT2 again shows the largest deviation, overestimating the contraction by 0.0214 {\AA}.
For the O 1s core-ionized state, EOM-DSRG-PT2 fails to capture the correct trend: while experiment shows a bond-length increase of 0.0370 {\AA}, it instead predicts a contraction of 0.0332 {\AA}.
The other methods correctly predict bond elongation, but MR-ADC(2)-X significantly overestimates the magnitude, predicting an increase of 0.0589 {\AA}, which is more than 1.5 times the experimental value.

\begin{table*}[!htb]
	\caption{Vibrational constants for K-edge core-ionized states of \ce{N2} and \ce{CO} computed using various EOM-DSRG methods [PT2 = DSRG-MRPT2, PT3 = DSRG-MRPT3, and LDSRG(2) = MR-LDSRG(2)], as well as GAS-DSRG [MR-LDSRG(2) for \ce{N2+} and DSRG-MRPT3 for \ce{CO+}] and MR-ADC(2)-X.
	Difference in equilibrium bond-length between the core-ionized and ground state ($\Delta r_e$), vibrational frequency ($\omega_e$), and anharmonic constant ($\omega_e \chi_e$).
	Reference experimental values taken from~refs~\citenum{matsumoto.2006.10.1016/j.cplett.2005.09.094} and \citenum{ueda.2009.10.1140/epjst/e2009-00978-7}.}
	\label{tab:spec}
	\setlength{\extrarowheight}{2pt}
	\setstretch{1}
	\centering
	\begin{tabular}{l*{6}{c}}
		\hline
		
		\hline
		 Parameters & GAS-DSRG & PT2 & PT3 & LDSRG(2) & MR-ADC(2)-X & Exp. \\ 
		\hline
		\multicolumn{7}{c}{\ce{N2}, N K-edge ($^2\Sigma_u^+$)} \\
		$\Delta r_e$ (\AA) & -0.0229 & -0.0350 & -0.0294 & -0.0298 & -0.0286 & -0.0240\\
		$\omega_e$ (cm$^{-1}$) & 2410 & 2581 & 2510 & 2508 & 2571 & 2407\\
		$\omega_e \chi_e$ (cm$^{-1}$) & 20 & 17 & 19 & 18 & 21 & -\\
		\hline
		\multicolumn{7}{c}{\ce{N2}, N K-edge ($^2\Sigma_g^+$)} \\
		$\Delta r_e$ (\AA) & -0.0191 & -0.0313 & -0.0256 & -0.0261 & -0.0254 & -0.0186\\
		$\omega_e$ (cm$^{-1}$) & 2411 & 2582 & 2509 & 2507 & 2568 & 2414\\
		$\omega_e \chi_e$ (cm$^{-1}$) & 21 & 18 & 20 & 19 & 22 & -\\
		\hline
		\multicolumn{7}{c}{\ce{CO}, C K-edge} \\		
		$\Delta r_e$ (\AA) & -0.0464 & -0.0728 & -0.0541 & -0.0546 & -0.0600 & -0.0514\\
		$\omega_e$ (cm$^{-1}$) & 2427 & 2611 & 2540 & 2554 & 2658 & 2479\\
		$\omega_e \chi_e$ (cm$^{-1}$) & 23 & 22 & 20 & 20 & 25 & 23\\
		\hline
		\multicolumn{7}{c}{\ce{CO}, O K-edge} \\		
		$\Delta r_e$ (\AA) & 0.0307 & -0.0332 & 0.0342 & 0.0337 & 0.0589 & 0.0370\\
		$\omega_e$ (cm$^{-1}$) & 1868 & 2210 & 1874 & 1885 & 1548 & 1864\\
		$\omega_e \chi_e$ (cm$^{-1}$) & 9 & 16 & 8 & 8 & 43 & 7\\
		\hline	
		
		\hline
	\end{tabular}
\end{table*}

\begin{figure*}[t]
\centering
\includegraphics[width = 6in]{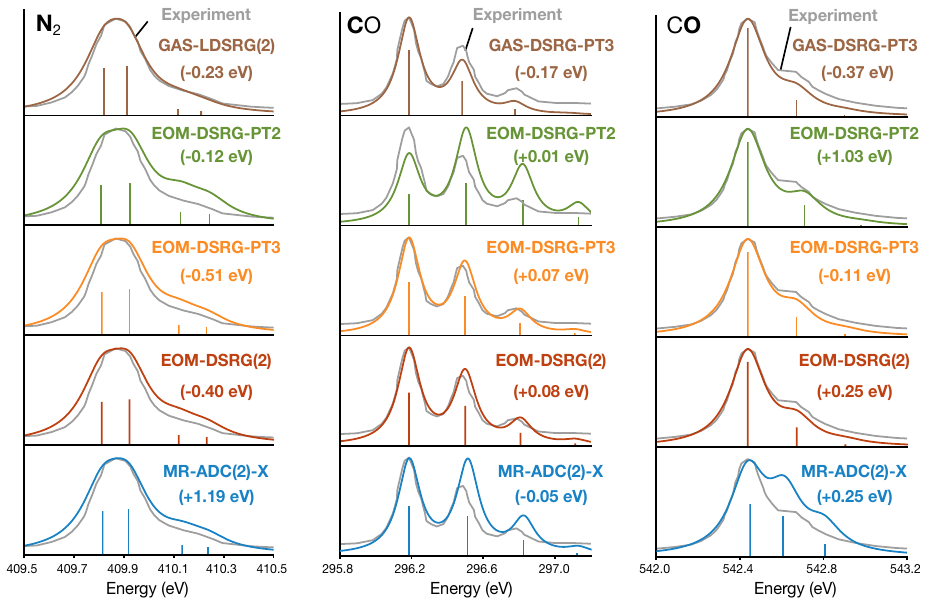} 
\caption{Vibrationally resolved X-ray photoelectron spectra (XPS) of (a) \ce{N2} N 1s, (b) \ce{CO} C 1s, and (c) \ce{CO} O 1s, simulated using various levels of EOM-DSRG theory, GAS-DSRG, and MR-ADC(2)-X. Experimental spectra are reproduced with permission from refs~\citenum{ueda.2009.10.1140/epjst/e2009-00978-7} and \citenum{matsumoto.2006.10.1016/j.cplett.2005.09.094}. For best comparison, all theoretical spectra are aligned to experiment by applying energy shifts, which are indicated in each panel. The Cohen-Fano interference effect is not included in the \ce{N2} simulations.\cite{ueda.2009.10.1140/epjst/e2009-00978-7}
\label{fig:spectra}
}
\end{figure*}

Lastly, we use the computed PECs to evaluate the vibrational levels of the core-ionized states and Franck--Condon factors. 
The vibrational eigenvalues and eigenstates are obtained using the discrete variable representation (DVR) method,\cite{lill.1982.10.1016/0009-26148283051-0, colbert.1992.10.1063/1.462100} with the potential energy values at each grid point evaluated through cubic spline interpolation.
\Cref{fig:spectra} presents both experimental and theoretical spectra for both \ce{N2} and \ce{CO}.
To enable direct comparison, energy shifts are applied to align the simulated spectra with the first experimental peak.
These shifts reflect the errors in the zero-point-corrected adiabatic transition energies.\cite{huang.2024.10.1021/acs.jctc.4c00835}
For the \ce{N2} system, all methods successfully reproduce the vibrational structure observed in the experiment.
Among these methods, EOM-DSRG-PT2 requires the smallest energy shift ($-0.12$~eV), likely due to fortuitous error cancellation.
GAS-LDSRG(2) requires a shift of $-0.23$~eV, while EOM-DSRG-PT3 and EOM-LDSRG(2) require slightly larger shifts of $-0.51$ and $-0.40$~eV, respectively.
For \ce{CO}, the GAS-DSRG-PT3, EOM-DSRG-PT3, and EOM-LDSRG(2) methods accurately capture the vibrational structure for both C and O core-ionized states.
In contrast, EOM-DSRG-PT2 fails to predict the correct relative intensities for the C 1s spectrum, and MR-ADC(2)-X fails for both C 1s and O 1s spectra.
Among all tested methods, EOM-DSRG-PT3 exhibits the best agreement with experiment for both C 1s and O 1s ionized states and performs comparably to EOM-LDSRG(2).

\subsection{X-ray photoelectron spectrum of ozone}
Our last application of the EOM-DSRG methods focuses on the more challenging problem of the ozone molecule (\ce{O3}).
In its ground state, the wave function of ozone is dominated by a closed-shell configuration featuring a doubly occupied $1a_2$ orbital and a smaller contribution from the doubly excited determinant $(1a_2)^{-2}(2b_1)^2$.\cite{miliordos.2014.10.1021/ja410726u}
The experimental XPS of ozone features two peaks that correspond to K-edge core ionization of the terminal O atoms ($\mathrm{O_T}$, resulting from the $2a_1$ and $1b_2$ orbitals, found at 541.5~eV) and the central atom ($\mathrm{O_C}$, $1a_1$ orbital, found at 546.2~eV).\cite{banna1977study}
Unlike valence ionization energies,\cite{Musial.2009.10.1063/1.3265770} EOM methods can struggle to accurately predict the core-excitation energies of \ce{O3} and their splitting ($\Delta_{\mathrm{CT}} \approx$ 4.70~eV).\cite{coriani.2015.10.1063/1.4935712,demoura.2022.10.1039/D1CP05476G, mazin.2023.10.1021/acs.jctc.3c00477}

\begin{figure}[!htb]
	\centering
	\includegraphics[width=3.125in]{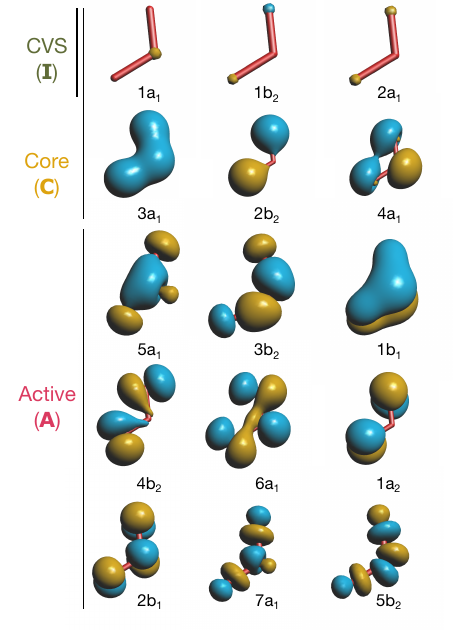}
	\caption{Semicanonical orbitals for the CVS, core, and active spaces of the ozone molecule computed using CASSCF with the (12e, 9o) active space.} 
	\label{fig:ozone_orbs}
\end{figure}

To simulate the XPS of ozone, we adopt the same active space used in a previous MR-ADC study,\cite{demoura.2022.10.1039/D1CP05476G} including 12 electrons in 9 active orbitals (four active electrons and three 2p orbitals from each oxygen atom).
The CASSCF semicanonical orbitals for the CVS, core, and active spaces are shown in \Cref{fig:ozone_orbs}.
All EOM-DSRG computations use the recontracted cc-pCVTZ-X2C basis and account for relativistic effects using the one-electron spin-free X2C (1e-sf-X2C) method.\cite{cfour_basis}
For the EOM-DSRG-PT3 approach, we also examine a truncated variant including only 1h excitations in the EOM excitation operator (\cref{eq:eom operator}), which we denote as EOM-DSRG-PT3-S.
SR-ADC, EOM-CCSD, and MR-ADC results are taken from the work of De Moura and Sokolov,\cite{demoura.2022.10.1039/D1CP05476G} while nonrelativistic single-reference EOM/LR-CC results were provided by Coriani.\cite{coriani2025private}

In \Cref{tab:ozone}, we show a comparison of a variety of theoretical and experimental results for the XPS of ozone.
For the terminal peaks, we report the average core-ionization energy for the $(1a_1)^{-1}$ and $(1b_2)^{-1}$ states since they typically differ by less than 0.01 eV.
In reporting these data, we also note if the core electrons were frozen in the ground state computation (when applicable) and if relativistic effects were included in the Hamiltonian.
Relative to nonrelativistic, all-electron computations, the 1e-sf-X2C treatment increases core-ionization energies by 0.37--0.39~eV\cite{demoura.2022.10.1039/D1CP05476G} and freezing ground-state core orbitals reduces them by 1.25--1.27~eV. 
However, since these shifts are uniform for the core-ionized states, neither changes $\Delta_{\mathrm{CT}}$ by more than 0.02~eV.

We first analyze the results for single-reference methods.
As shown in \Cref{tab:ozone}, the ADC(2) and ADC(2)-X core-ionization energies are within 0.7~eV from experiment; going to the next order, SR-ADC(3) worsens the core-ionization energies, overestimating them by 5--7~eV, suggesting cancellation of errors at second order or poor convergence behavior of the ADC series at higher order.
The EOM-CC results based on the all-electron CVS scheme\cite{coriani.2015.10.1063/1.4935712} overestimate the ionization energies at the CCSD level by up to 3~eV. The addition of triples via the CCSDR(3)\cite{christiansen.1996.10.1063/1.472007, coriani.2012.10.1103/PhysRevA.85.022507, coriani.2012.10.1021/ct200919e} and CC3\cite{christiansen.1995.10.1063/1.470315, koch.1997.10.1063/1.473322, myhre.2016.10.1063/1.4959373, myhre.2018.10.1063/1.5011148, myhre.2019.10.1021/acs.jpca.9b06590} methods reduces the error down to less than 0.8~eV (CC3).
Nevertheless, across the EOM-CC hierarchy, the value of $\Delta_{\mathrm{CT}}$ is consistently predicted to be in a narrow range (5.1--5.4~eV).

Among the multireference ADC methods, MR-ADC(2)-X best reproduces the ionization energies, with errors of only 0.50--0.77~eV and $\Delta_{\mathrm{CT}}$ = 4.43~eV.
For the EOM-DSRG methods, the most accurate LDSRG(2) treatment yields a value of $\Delta_{\mathrm{CT}}$ = 4.01~eV, with absolute excitation energies overestimated by 3.1 and 2.5~eV for the $\mathrm{O_T}$ and $\mathrm{O_C}$ transitions, respectively.
A comparison of the regular EOM-DSRG-PT3 with the 1h-only variant (-S) shows that the missing orbital relaxation and coupling to 2h1p configurations lead to severely overestimating the core-ionization energies by more than 20~eV.
In contrast to the EOM methods, approaches such as state-averaged DSRG-MRPT2/3 and multistate RASPT2\cite{tenorio.2022.10.1039/D2CP03709B} optimize variational parameters specifically for the target states.
As a consequence, these methods yield more consistent results and accurately reproduce the experimental core-ionization energy and $\Delta_{\mathrm{CT}}$.
Overall, this example shows that for both single and multireference variants of the EOM methods, it is essential to include higher-order dynamical correlation effects (3h2p) to accurately reproduce the core-ionization energies.

\begin{table*}[!htb]
	\caption{Theoretical and experimental O K-edge core-ionization energies (in eV) and spectroscopic factors for the terminal ($\mathrm{O_T}$) and central ($\mathrm{O_C}$) oxygen atoms. The energy splitting (in eV) between the $\mathrm{O_T}$ and $\mathrm{O_C}$ peaks is reported as $\Delta_{\mathrm{CT}}$. EOM-DSRG methods are labeled as PT2, PT3, PT3-S, and LDSRG(2), corresponding to EOM-DSRG-PT2, EOM-DSRG-PT3, EOM-DSRG-PT3-S, and EOM-LDSRG(2), respectively. All EOM-DSRG results employ a value of $s = 0.5\ E_{\mathrm{h}}^{-2}$, while SA-DSRG results use $s = 1.0\ E_{\mathrm{h}}^{-2}$.}
	\label{tab:ozone}
	\setlength{\extrarowheight}{2pt}
	\setstretch{1.1}
	\centering
	\small
	\begin{threeparttable}
	\begin{tabular}{lccllc}
		\hline
		
		\hline
	    Method & Frozen core$^a$ & 1e-sf-X2C$^b$ & $\mathrm{O_T}$ (2$\mathrm{a_1^{-1}}$/1$\mathrm{b_2^{-1}}$) & $\mathrm{O_C}$ (1$\mathrm{a_1^{-1}}$)& $\Delta_{\mathrm{CT}}$ \\
	    \hline
	    SR-ADC(2)$^c$ & No & Yes & 541.02 (1.461) & 546.83 (1.490) & 5.81 \\
	    SR-ADC(2)-X$^c$ & No & Yes & 540.87 (1.471) & 546.46 (1.492) & 5.59 \\
	    SR-ADC(3)$^c$ & No & Yes & 548.55 (1.635) & 551.75 (1.664)  & 3.20 \\
	    EOM-CCSD$^c$ & No & No & 544.15  &  549.23 & 5.07 \\
	    EOM-CCSD$^d$ & Yes & No & 542.90 & 547.96 & 5.06 \\
	    CCSDR(3)$^d$ & No & No & 542.36 & 547.43 & 5.07 \\
	    CC3$^d$ & No & No & 540.78 & 546.16 & 5.38 \\
	    MR-ADC(2)$^c$ & No & Yes & 543.85 (1.531) & 548.48 (1.552) & 4.63 \\
	    MR-ADC(2)-X$^c$ & No & Yes & 541.00 (1.291) & 545.43 (1.380) & 4.43 \\
	    EOM-DSRG-PT2 & Yes & Yes & 545.49 (1.476) & 549.17 (1.488) & 3.68 \\
	    EOM-DSRG-PT3 & Yes & Yes & 544.66 (1.431) & 548.81 (1.473) & 4.15 \\
	    EOM-DSRG-PT3 & No & Yes & 545.93 (1.432) & 550.08 (1.475) & 4.15 \\
	    EOM-DSRG-PT3-S & Yes & Yes & 563.33 (2.000) & 567.75 (2.000) & 4.42 \\
	    EOM-LDSRG(2) & Yes & Yes & 544.64 (1.435) & 548.65 (1.472) & 4.01 \\
	    SA-DSRG-MRPT2$^e$ & No & Yes & 540.84 & 545.53 & 4.69 \\
	    SA-DSRG-MRPT3$^e$ & No & Yes & 541.35 & 545.64 & 4.29 \\    
	    MS-RASPT2$^f$ & No & No & 541.52 & 547.25 & 5.73 \\	    
	    Experiment$^g$ & - & - & 541.5 &  546.2 & 4.70 \\
		\hline
			
		\hline
	\end{tabular}
	\end{threeparttable}\\
	$^a$ Yes = core electrons are frozen in the ground-state computation, No = all-electron ground-state computation.
	$^b$ Yes = one-electron 1e-sf-X2C treatment and cc-pCVTZ-X2C basis, No = nonrelativistic Hamiltonian and cc-pCVTZ basis.
	$^c$ From ref \citenum{demoura.2022.10.1039/D1CP05476G}. $^d$ From ref \citenum{coriani2025private}. $^e$ From ref \citenum{huang.2024.10.1021/acs.jctc.4c00835}, with cc-pVQZ basis set. $^f$ From ref \citenum{tenorio.2022.10.1039/D2CP03709B}, with cc-pVTZ basis set. $^g$ From ref \citenum{banna1977study}.
\end{table*}

\section{Conclusion}
\label{sec:conclusion}
In this work, we formulate and implement a core-valence separated multireference equation-of-motion driven similarity renormalization group method (CVS-IP-EOM-DSRG) and apply it to compute core-ionization energies and simulate X-ray photoelectron spectra (XPS).
To ensure rigorous core intensivity of the ionization energies, we propose a modification of the ground-state DSRG formalism that solves projective equations for core-virtual singles amplitudes.

The CVS-EOM-DSRG framework is tested using a ground state obtained at three truncation levels: the perturbative DSRG-MRPT2 and DSRG-MRPT3 methods, and the non-perturbative MR-LDSRG(2) approach.
For each truncation level, we systematically investigate the dependence on the flow parameter and the choice of the CVS scheme.
We find that ionization energies show only a weak dependence on the flow parameter within the recommended range.
Two CVS schemes are compared, differing in their treatment of the occupied orbitals in the MR-DSRG ground-state wavefunction.
We observe that applying the frozen-core approximation to the ground state and applying restrictions consistent with the CVS to the EOM operator systematically lowers the computed ionization energies, leading to very good agreement with benchmark CVS-EOMIP-CCSDTQ results.
Numerical tests show that enforcing a subset of the projective conditions restores core intensivity of the ionization energies while introducing only negligible differences in the ground-state energy.
We then benchmark the CVS-IP-EOM-DSRG method based on the three ground state truncation schemes by computing K-edge vertical ionization energies for a set of small molecules.
Our results show that the DSRG-PT3 truncation scheme balances well computational cost and accuracy: it slightly underperforms compared to state-specific methods like GAS-DSRG and CVS-EOM-CCSDT (due to the lack of orbital relaxation and higher-order excitations). Still, it is on par or outperforms the accuracy of the CVS-MR-ADC(2) and CVS-MR-ADC(2)-X schemes.
Moreover, the PT3 truncation level matches the accuracy of CVS-IP-EOM-LDSRG(2), while offering improved computational efficiency by avoiding iterative ground-state optimization and its associated convergence challenges.

To assess the applicability of these methods to molecules with open-shell character, we compute potential energy curves for the N K-edge ionized states of \ce{N2} and the C and O K-edges ionized states of \ce{CO}, comparing them with CVS-MR-ADC(2)-X results using GAS-DSRG curves as references.
While all methods yield qualitatively correct potentials, CVS-IP-EOM-DSRG-PT3 and -LDSRG(2) accurately predict vibrational constants and show small nonparallelism errors.
Using these curves, we simulate the vibrationally resolved XPS of \ce{N2} and \ce{CO}, finding that the DSRG-PT3 and LDSRG(2) truncation schemes show excellent agreement with experiment. At the same time, CVS-IP-EOM-DSRG-PT2 and CVS-MR-ADC(2)-X fail to capture correct vibrational intensities in some cases.
We also applied the CVS-EOM-DSRG method to compute the XPS of ozone, which is more challenging for EOM methods that lack 3h2p excitations.
Both CVS-EOM-DSRG-PT3 and CVS-EOM-LDSRG(2) correctly reproduce the energy splitting between ionization energies to within 0.7~eV, although a sizable shift ($\approx$3.1 eV) is needed to match the lowest experimental ionization energy.

Overall, the relative accuracy of the new methods introduced follows the trend: CVS-EOM-LDSRG(2) $\approx$ CVS-EOM-DSRG-PT3 $>$ CVS-EOM-DSRG-PT2.
In particular, CVS-EOM-DSRG-PT3 offers a good balance between accuracy and computational efficiency, making it a reliable choice for modeling core-ionized states.
Extending the method to incorporate particle-preserving excitation operators will enable applications to X-ray absorption (XAS) and UV/vis spectroscopy, and efforts in this direction are currently ongoing in our group.
Finally, although numerical tests in this study indicate that enforcing projective conditions for core-virtual singles has little effect on the MR-DSRG ground-state energy, a systematic investigation is needed to better establish the impact of this choice on the accuracy and numerical stability of ground-state computations.

\section*{Author Information}
Shuhang Li (ORCID: 0000-0002-1488-9897)\\
Zijun Zhao (ORCID: 0000-0003-0399-2968)\\
Francesco A. Evangelista (ORCID: 0000-0002-7917-6652)

\section*{Supporting Information}
See the Supporting Information for additional details: 1) The choice of active spaces and reference energies for the CASSCF computations, where applicable.
2) Vertical ionization energies and spectroscopic factors of ozone.
3) A zip file containing comma-separated value (CSV) files of the raw data for all figures in the main text.

\section*{Acknowledgements}
We thank Sonia Coriani for providing the EOM-CCSD, CCSDR(3), and CC3 results for ozone reported in the article.
We thank Kevin Marin and Meng Huang for insightful discussions.
This research was supported by the U.S. National Science Foundation under award number CHE-2312105.
S.L. was supported to develop the \textsc{NiuPy} package by the National Science Foundation and the Molecular Sciences Software Institute under Grant No. CHE-2136142.

\section*{Data availability}
All data are available upon reasonable request.
The software used to produce the data presented in this work is available in an accompanying public code repository.\cite{Li_niupy_Open-source_implementation_2025}

\bibliography{ref}
\clearpage
\includepdf[pages=-]{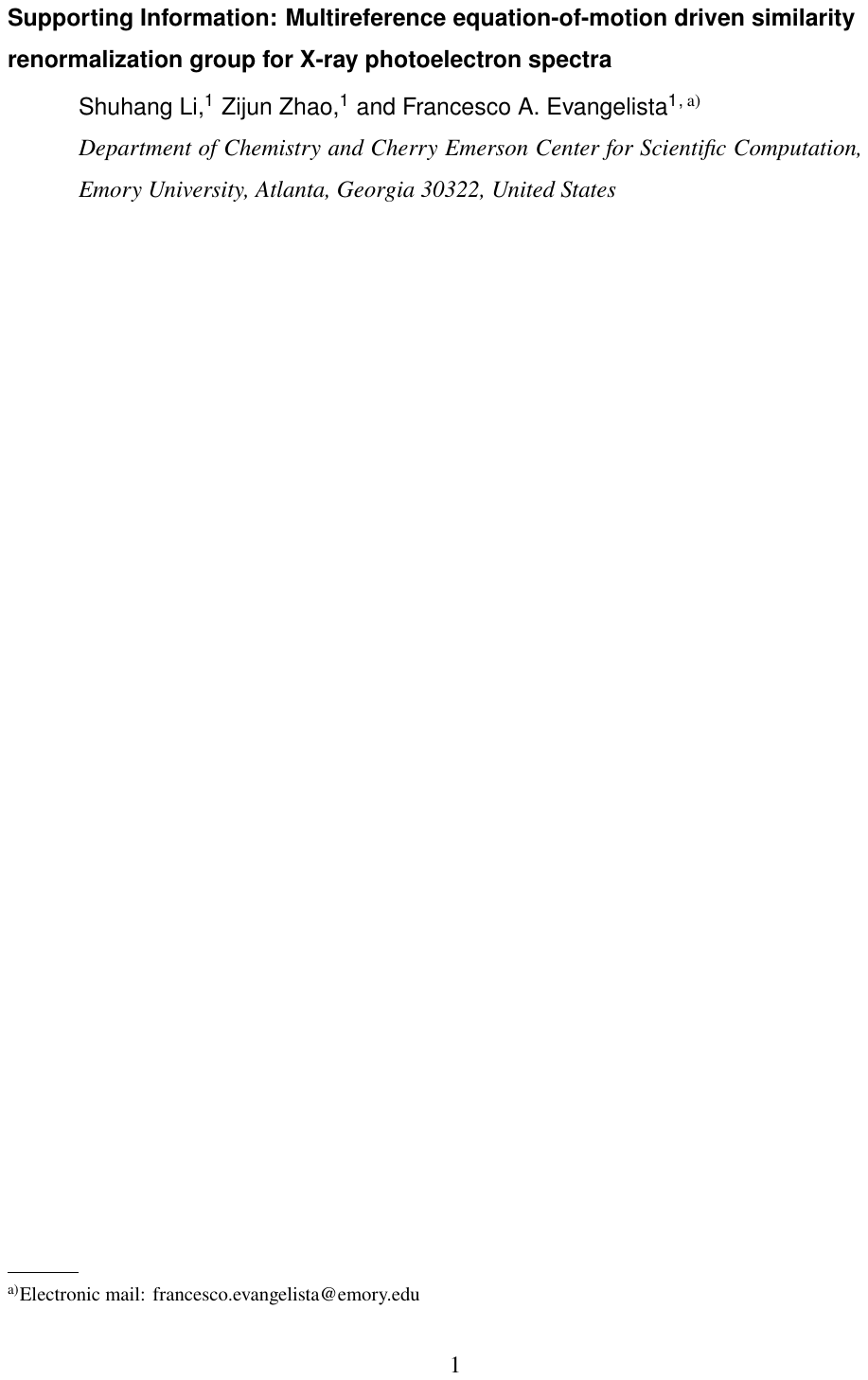}

\end{document}